\documentclass[pre, aps,showpacs, notitlepage, groupedaddress, superscriptaddress, twocolumn, numerical]{revtex4-1}

\usepackage[utf8x]{inputenc}
\usepackage{color}
\usepackage{bbm}

\usepackage{nicefrac}
\usepackage{amsfonts,amsmath,amssymb,stmaryrd}
\usepackage{braket}

 \usepackage[autostyle=true]{csquotes}

\usepackage{tabularx}
\usepackage{multirow} 
\usepackage{hhline}
\usepackage{graphicx}
\usepackage{subfigure}  
\usepackage{bbm} 
\usepackage{hyperref}
\usepackage[pdftex]{epsfig}
\usepackage{mathrsfs}
\usepackage{verbatim}
\usepackage{centernot}
\usepackage{ulem}
\usepackage{array}
\usepackage{cancel}
\usepackage{ifthen}
\usepackage{bm}	
\usepackage{siunitx}
\usepackage{todonotes}
\usepackage{setspace}
\usepackage{bm} 

\InputIfFileExists{gitinfo-latex.inc}{}{}

\usepackage{listings}
\usepackage{color}
\usepackage{ulem}

\usepackage[acronym,nomain]{glossaries}

\newcommand\subplotlabel{}
\def\subplotlabel[#1](#2)(#3){%
	\node[fill=white, fill opacity=0.6, text opacity=1,  minimum size=5mm,inner sep=0pt,outer sep=0pt, #1] at (#2) {{\large \textsf{\textbf{#3}}}};		
}

\AtBeginDocument{\renewcommand{\d}{\mathrm{d}}}

\newcommand{\Ti}{T_{0}}
\newcommand{\Tc}{T_\text{R}}

\newcommand{\ramancooling}{DRSC{}}

\newcommand{\tempFOVEinit}{\SI{12.1 \pm 1.1}{\micro K}}

\newcommand{\tempFOVEcooled}{\SI{2.9 \pm 0.2}{\micro K}}

\begin{document}

\title{{Nonequilibrium thermodynamics and optimal cooling of a dilute atomic gas}}
\author{Daniel Mayer}
	\affiliation{Department of Physics and Research Center OPTIMAS, Technische Universit\"at Kaiserslautern, Germany}

	\author{Felix Schmidt}
	\affiliation{Department of Physics and Research Center OPTIMAS, Technische Universit\"at Kaiserslautern, Germany}
	
	\author{Steve Haupt}
	\affiliation{Department of Physics and Research Center OPTIMAS, Technische Universit\"at Kaiserslautern, Germany}
	
	\author{Quentin Bouton}
	\affiliation{Department of Physics and Research Center OPTIMAS, Technische Universit\"at Kaiserslautern, Germany}

	\author{Daniel Adam}
	\affiliation{Department of Physics and Research Center OPTIMAS, Technische Universit\"at Kaiserslautern, Germany}
	
	\author{Tobias Lausch}
	\affiliation{Department of Physics and Research Center OPTIMAS, Technische Universit\"at Kaiserslautern, Germany}
	
	\author{Eric Lutz}
	\affiliation{Institute for Theoretical Physics I, University of Stuttgart, D-70550 Stuttgart, Germany}
	
	\author{Artur Widera}
	
	\affiliation{Department of Physics and Research Center OPTIMAS, Technische Universit\"at Kaiserslautern, Germany}
	\affiliation{Graduate School Materials Science in Mainz, Gottlieb-Daimler-Strasse 47, 67663 Kaiserslautern, Germany}

\begin{abstract}
{Characterizing and optimizing thermodynamic processes far from equilibrium is a challenge. This is especially true for nanoscopic systems made of few particles. We here {theoretically and} experimentally investigate the nonequilibrium dynamics   of a gas of few noninteracting Cesium atoms confined in a nonharmonic optical dipole trap and exposed to degenerate Raman sideband cooling pulses. We determine the axial phase-space distribution of the atoms after each Raman cooling pulse by tracing the evolution of the gas with position-resolved fluorescence imaging. We evaluate from it the entropy production and the statistical length between each cooling steps. 
{A single  Raman pulse leads to a nonequilibrium state that does not thermalize on its own, due to the absence of  interparticle collisions.} Thermalization may be achieved by combining free phase-space evolution and trains of cooling pulses.  We  minimize the  entropy production to a target thermal state to specify the optimal spacing between a sequence of equally spaced pulses and  achieve in this way optimal thermalization. We finally use the statistical length to verify a refined version of the second law of thermodynamics {Altogether, these findings provide a general, theoretical and experimental, framework to analyze and optimize far-from-equilibrium processes of few-particle systems.} } \end{abstract}
\maketitle

\section{Introduction}

{Nonequilibrium processes are omnipresent in nature. Owing to their complexity and diversity, their description far away  from thermal equilibrium is nontrivial \cite{leb08}. A defining property of out-of-equilibrium systems is that they dissipate energy in the form of heat, leading to an irreversible increase of their entropy. The irreversible entropy production is thus a central quantity of nonequilibrium thermodynamics, the same way that entropy is a central quantity of equilibrium physics \cite{leb08}. In the past decades, the laws of thermodynamics have been successfully extended to small nonequilibrium systems \cite{bus05,sek10,sei12,jar11,cil13}. In these systems, thermal fluctuations can no longer be neglected and thermodynamic variables are therefore random. In particular, the second law has been generalized in the form of fluctuation theorems that quantify the occurrence of negative entropy production events \cite{bus05,sek10,sei12,jar11,cil13}. The stochastic properties of the nonequilibrium entropy production have been extensively investigated, both theoretically and experimentally, for microscopic systems such as colloidal particles \cite{bus05,sek10,sei12,jar11,cil13,wan02,car04,sch05,bli06,dou06,tie06}.  On the other hand, only few experiments {have probed nonequilibrium thermodynamics in} {nanoscopic} systems so far. These include one-particle systems, such as a single  spin-1/2 \cite{bat14,bat15} or a single harmonic oscillator \cite{an15,bru18}, two-spin systems \cite{mic19,pal19}, and many-particle systems, such as cold-atomic gases \cite{bru18,kin06,gri12,cer17}.  {However, to our knowledge, no such nonequilibrium thermodynamic experiment has been realized in the intermediate regime of few-particle systems.}

    {The laws of thermodynamics are not only of fundamental but also of practical importance. A primary objective of thermodynamics is thus to  optimize processes.} Optimization goals vary depending on the  application, ranging from 
 the minimization of dissipation to  the maximization of work output or of  cooling power \cite{bej06}.
For macroscopic systems, the properties of optimal transformations have been studied  within finite-time thermodynamics \cite{sal83,and84,nul85,and11}.
The two central quantities of this approach are the  entropy production, that characterizes energy dissipation, and the thermodynamic length, that measures the distance from equilibrium at which a system operates.
Both are commonly calculated in the linear response regime by expanding thermodynamic potentials, such as  entropy or  internal energy, to {second} order  around  equilibrium  \cite{sal83,and84,nul85,and11}. {Optimization schemes are usually developed by minimizing one of the two}.
These techniques have been employed to optimize fractional distillation and other  processes \cite{sal83,and84,nul85,and11,sal98,sal01,nul02,che99}.
On the other hand, for microscopic systems, where thermal fluctuations are sizable, this optimization framework has been extended to the level of single trajectories within stochastic thermodynamics for linear \cite{sch07} and nonlinear \cite{aur11} systems.
Methods to theoretically compute and experimentally evaluate the thermodynamic length have been proposed \cite{cro07,fen09,siv12,gin16}. {However, despite these theoretical studies, such nonequilibrium optimization schemes have still to be demonstrated experimentally. In particular,  thermodynamic distances have not been measured yet.} 

{A further complication arises in atomic systems.} A central assumption of  finite-time  thermodynamics and stochastic thermodynamics is {indeed} that systems  are coupled to ideal heat baths that induce full phase-space thermalization, {that is, of both position and momentum degrees of freedom.}
However, this hypothesis is often not fulfilled at the atomic level.
A prominent instance is provided by laser cooling  of atoms which plays an essential role in the study of new states of matter and high-resolution spectroscopy \cite{coh11}.
Most laser cooling schemes  only induce  thermalization of the momentum degrees of freedom \cite{met99}.
In dense atomic samples, frequent atomic collisions redistribute the energy and establish thermal equilibrium.
By contrast,  in dilute  gases with  rare interparticle collisions, these nonideal reservoirs lead  to far from equilibrium  states {that do not thermalize on their own}.
Their description thus lies  outside the currently existing  framework.
New experimental and  theoretical tools are hence required to achieve their thermalization.

\begin{figure}
	\begin{center}
		\includegraphics[scale=1]{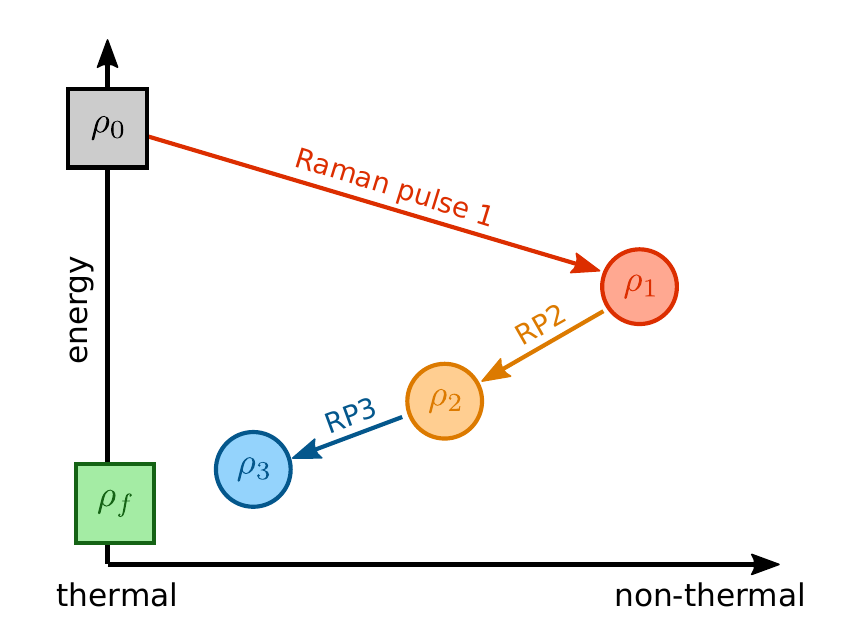}
	\end{center}
	\caption{An initial thermal state $\rho_0$ of a noninteracting gas of Cs atoms at temperature $\Ti$ is cooled and rethermalized towards a final state $\rho_f$  at the lower (Raman) temperature $\Tc$.
	This is achieved by applying a train of equally spaced  degenerate Raman sideband cooling  pulses (red, orange and blue), which only thermalize the momentum degree of freedom. The first cooling pulse thus creates a nonthermal state while the successive Raman pulses, {combined with free  evolution}, drive the system towards the target thermal state $\rho_f$.	} 
	\label{fig:overview}
\end{figure}

{We here report the theoretical and experimental investigation of} the nonequilibrium dynamics and the thermalization of a dilute gas of Cesium atoms confined in an optical dipole trap \cite{met99}, and illuminated by laser pulses for degenerate Raman sideband cooling  (DRSC) \cite{vul98,ker00,han00}.}  This technique  is a standard subrecoil cooling scheme for a variety of atomic systems  \cite{tre01,web03,mon95,des04,lee96,che15,par15,gro17,kau12,tho13}.
The present study of a few-particle system coupled to an engineered bath allows us to experimentally access  key nonequilibrium quantities in a well-controlled atomic setup. {It further gives us the opportunity to illustrate and validate our general nonequilibrium optimization approach with a common laser cooling example.} {We determine {in particular}, for each thermalization step, the nonequilibrium entropy production and  a generalized thermodynamic length appropriate for these nonideal reservoirs. We use the former  quantity to optimize the cooling of the few-atom system and the latter one to gain physical insight into the optimal cooling process and verify a refined version of the second law of thermodynamics known as the horse-carrot theorem \cite{and11,sal98}.} 

In our experiment, short pulses of {Raman cooling} lasers are applied to an initially thermal cloud along the axial direction of our  nonharmonic trap. Axial and radial directions are  only weakly coupled, {rendering} the problem  essentially one-dimensional. 
The Raman pulses thermalize the atomic momentum {distribution} to the Raman temperature, {thus cooling the system,}  but leave the position distribution unchanged. {They hence create for most initial conditions} a nonequilibrium state  that does not thermalize on its own,  due to the absence of interparticle collisions. 
In order to realize complete phase-space thermalization {at the Raman temperature}, we devise protocols consisting of a train of Raman pulses separated by intervals of free evolution (Fig.~\ref{fig:overview}).
For concreteness, we consider a sequence of three {equally spaced} pulses. The first Raman pulse (RP1) decreases the energy of the gas and moves it out of equilibrium. 
The second and third pulses (RP2 and RP3) drive the gas back towards a thermal state while cooling it further.
{For quasi harmonic trapping potentials, thermalization is routinely established by using a pulse spacing of a quarter of the trap period~\cite{dep99}. {This method has, for instance, recently  led to the all-optical Bose-condensation of Rb atoms without an evaporative cooling stage \cite{hu17}}. For the strongly anharmonic potential {of our experiment}, the trap leads to nontrivial dynamics of the nonequilibrium states and raises the question of the choice of the pulse spacing in this case.
We seek the optimal pulse spacing $\tau$ by minimizing the entropic distance to the equilibrium target state $\rho_f$ at the Raman temperature, {employing both a static and a dynamical criterion, which lead to the same result. An analysis of the nonequilibrium statistical length furthermore reveals that optimal thermalization is mainly reached  during the first two cooling stages, with nearly equal statistical distances. 
}

{The outline of the paper is as follows. We begin in Sect.~II by deriving the nonequilibrium entropy production and the statistical length for the nonideal reservoirs occurring in the experiment. We further present the horse-carrot theorem and the two criteria used to optimize the thermalization.  In Sect.~III we illustrate the physical meaning of the static and dynamical optimization criteria  for the analytically solvable case of a harmonic trapping potential. We additionally present the experimental setup in Sect.~IV and the  numerical phase-space reconstruction of the phase-space distributions in Sect.~V. Finally, in Sect.~VI, we  demonstrate optimal thermalization in a strongly nonharmonic trap and an experimental verification of the horse-carrot theorem.}

{\section{Nonequilibrium quantities and optimization criteria}}

As illustrated in Fig.~\ref{fig:overview}, the goal of the DRSC protocol is to reach the final thermal state
\begin{align}
	\rho_f(z, p_z) \propto \exp\left( -\frac{V(z)}{k \Tc} - \frac{p_z^2}{2mk\Tc} \right)
\end{align}
at the DRSC temperature $\Tc$, where $\rho(z, p_z)$ is the projected phase-space density onto the $(z,p_z)$ plane, which is the relevant subspace for our experiment, {and $V(z)$ denotes the axial potential}.
In order to quantify the approach of a nonthermal state $\rho_i$  produced by the DRSC scheme to the final target state $\rho_f$, we employ the relative entropy between these two states~\cite{cov06}
\begin{align}
	D(\rho_i||\rho_f) = \int dz dp_z \ \rho_i(z, p_z) \ln \left( \frac{\rho_i(z, p_z)}{\rho_f(z, p_z)} \right). 
\end{align}
Similarly, the corresponding quantities $D(f_i||f_f) = \int dz\, f_i \ln(f_i/f_f)$ and $D(\tilde{f}_i||\tilde{f}_f) = \int dp_z\, \tilde{f}_i \ln(\tilde{f}_i/\tilde{f}_f)$ can be defined for the respective position and momentum projections, $f(z)$ and $\tilde{f}(p_z)$, of the phase-space distribution.
The relative entropy is an information-theoretic quantity that  satisfies the important  property that $D(\rho_i||\rho_f) \geq 0$, equality being only achieved  for $ \rho_i = \rho_f$ ~\cite{cov06}.
This renders the relative entropy a useful indicator for the approach to the final target state.
}

The relative entropy also possesses a simple thermodynamic interpretation \cite{pro76,sch80,esp10,def11}.
For a nonequilibrium process from an initial thermal state $\rho_0$, at inverse temperature $\beta_0=(k T_0)^{-1}$, to a final thermal state $\rho_f$, at inverse temperature $\beta_f=(k T_f)^{-1}$, the (axial) Gibbs-Shannon entropy, $S= - \int dzdp_z \,\rho \ln\rho$  satisfies \cite{pro76,sch80,esp10,def11}
\begin{align}
\Delta S =S_f-S_0= \beta_f Q + \Sigma. 
\end{align}
Here, $Q= \int dzdp_z (\rho_f-\rho_0) H$ is the heat absorbed by the system, $H$ its Hamiltonian  and $\Sigma =  D(\rho_0||\rho_f)$ the nonequilibrium entropy production given as the relative entropy between initial and final states.

For a discrete sequence of nonthermal intermediate states $\rho_i$, {$(i=1, 2, 3)$}, as created after each {Raman cooling}  pulse in our experiment, the entropy production associated with each step  reads 
{$\Sigma_i = D\left(\rho_{i-1}||\rho_f\right) - D\left(\rho_{i}||\rho_f\right)$ (Appendix A).
	The statistical length defined as,
	\begin{equation}
	\label{1}
	L_i =\sqrt{2 \Sigma_i} = \sqrt{2[D\left(\rho_{i-1}||\rho_f\right) - D\left(\rho_{i}||\rho_f\right)]}, 
	\end{equation}
	{then quantifies the  distance from equilibrium at which the system operates. It vanishes when $\rho_i = \rho_f$ for all $i$}}. Equation \eqref{1} reduces to the usual thermodynamic length in the limit of quasistatic processes where all the intermediate states are close to thermal \cite{nul85,and11,sal98,sal01,nul02}.
	{The above nonequilibrium quantities allow the investigation of not only the final state reached after the application of the DRSC protocol but also of the cooling process itself, by providing direct information on the intermediate states}.
	
	{The total entropy production $\Sigma = \sum_{i=1}^n \Sigma_i$ (multiplied by $T_f$) is a measure of the amount of energy that is irreversibly extracted from the system during thermalization \cite{pro76,sch80,esp10,def11}. It  is bounded from below by the square of the total statistical length $L= \sum_{i=1}^n L_i$ divided by twice the number of steps (Appendix A)
	\begin{equation}
	 \Sigma \geq \frac{L^2}{2n}
	 \end{equation}in analogy to the horse-carrot theorem \cite{and11,sal98,sal01}. {The name horse-carrot process finds its origin in the  analogy with a system (the horse) which  is coaxed along a sequence of states by controlling  the state of its environment (the carrot). The importance of the horse-carrot theorem stems from the fact that it provides a sharper lower bound to the nonequilibrium entropy production than the second law  of thermodynamics which only requires $\Sigma \geq0$. It additionally implies that optimal quasistatic horse-carrot processes (for which inequality is replaced by an equality) correspond to steps of equal thermodynamic length \cite{sal83,and84,nul85,and11,sal98,sal01,nul02}. We shall find that this also holds exactly for a harmonic confining potential  and approximately for a nonharmonic trap for the generalized nonequilibrium statistical length (4) (Sect.~VI)}.

Commonly considered optimization schemes minimize the nonequilibrium entropy production with fixed initial and final states \cite{nul85,and11,sal98,sal01,nul02}. By contrast, {the state $\rho_i$ produced by the DRSC protocol} depends on the entire cooling sequence and is hence not fixed. Our strategy is therefore to minimize the entropic distance to the target thermal state $\rho_f$ and identify the final temperature with the Raman temperature,  $\beta_f=\beta_\mathrm{R}$. We concretely consider two optimization criteria:\\  

\noindent (1) Static criterion: the first condition minimizes the relative entropy  between $\rho_i$ and the target state $\rho_f$, $D\left(\rho_{i}||\rho_f\right)$. {This corresponds to {minimizing}  the entropy production $\Sigma_i$  [Eq.~\eqref{1}] in the case of successful thermalization}.\\

\noindent(2) Dynamical criterion: the second condition minimizes the amplitude of oscillations of the {positional} relative entropy, $\Delta D= \max_t D(f_i(t)||f_f) - \min_t D(f_i(t)||f_f)$,  during the free time evolution of the  atomic cloud after the Raman pulse. This criterion is based on the stationarity of a thermal state: for an equilibrium state, the distribution $f_i$ is constant in time and hence $\Delta D=0$. The closer the state is to equilibrium, the smaller the oscillation amplitude $\Delta D$.\\

The application of both optimization criteria requires to extract the  relative entropy $D$  from measured data.
\\

\section{Harmonic Case}

{In order to better understand the physical meaning of the above optimization criteria, we first consider the problem of a harmonic potential which is analytically solvable. In this case, the optimal pulse spacing is given by a quarter of the oscillation period \cite{dep99,hu17}. This result is intuitively clear as it corresponds to  the time needed to switch position and momentum axes in phase-space during free evolution.  Phase-space compression, and hence cooling and thermalization, is therefore optimal.}

\begin{figure}[t]
	\includegraphics[scale=1]{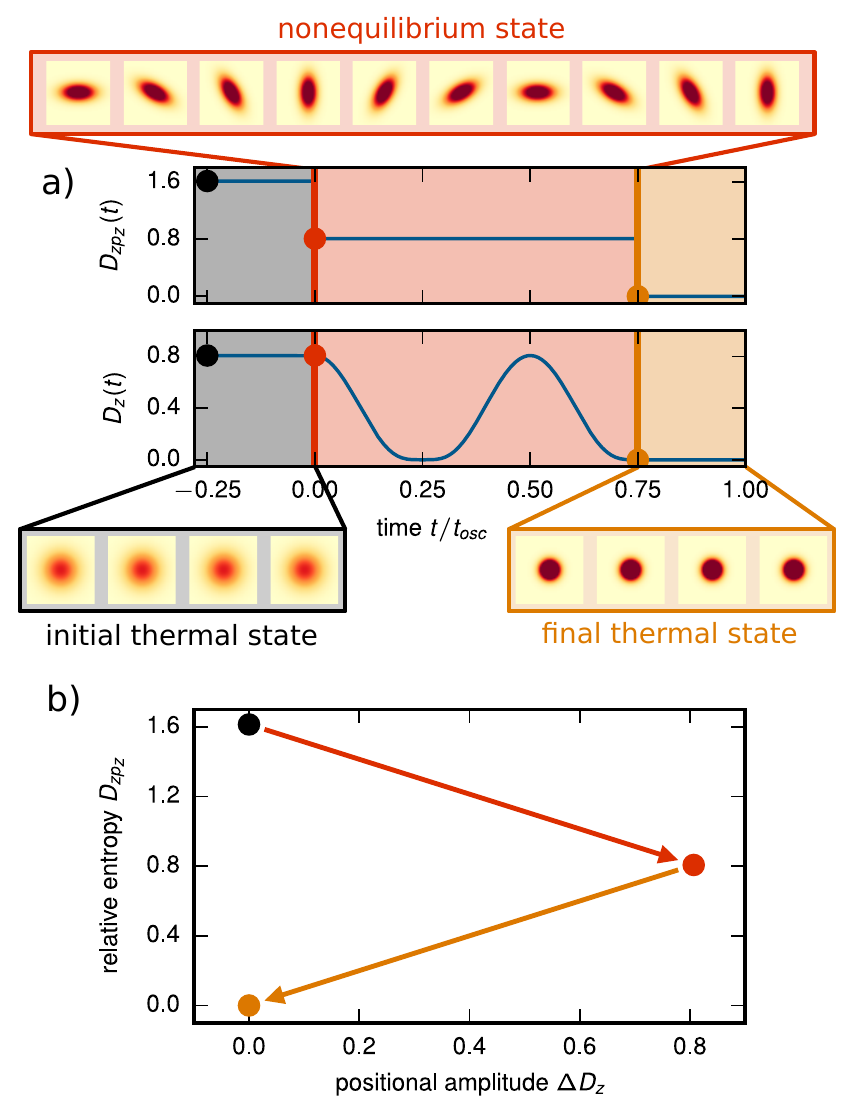}
	\caption{Illustration of the optimization criteria for the analytically solvable harmonic trap.
		For simplicity, a DRSC protocol with only two pulses is considered.
		\textbf{a)} Time evolution of the full and positional relative entropies, $D_{zp_z}(t)$ and $D_{z}(t)$.
		The vertical blue lines indicate the cooling pulses 
		applied at $t=0$ and $t=0.75\theta$.
		The colored background illustrates the different stages of the cooling protocol.
		Dots highlight the initial values (black), as well as values right after the first (red) and second (orange) DRSC pulse .
		\textbf{b)} Map of the cooling process, showing the full relative entropy $D_{zp_z}(t)$ at the points indicated in a) over the   amplitude $\Delta D_z(t)$ in position.
		The arrows indicate the impact of the two Raman pulses.}
	\label{fig:harmonic_case}
\end{figure}

{We analyze the phase-space dynamics by solving the Boltzmann equation for the  density $\rho(z, p_z,t)$~\cite{hua87}}
\begin{align}
\left( \frac{\partial}{\partial t} + \frac{p_z}{m} \frac{\partial}{\partial z} + {F(z)} \frac{\partial}{\partial p_z}  \right) \rho(z, p_z, t) = I_\mathrm{coll}(\rho, p_z), \label{eq:bme}
\end{align}
where $m$ is the atomic mass, $F(z)= -m\omega_z^2 x$  the force acting on the atom and $I_\mathrm{coll}$ the collision integral which takes in to account atomic interactions.
{For the few-atom samples that we consider}, atomic interactions are negligible and hence $I_\mathrm{coll} = 0$.
{Equation \eqref{eq:bme} can then  be solved exactly with the Gaussian ansatz  \cite{gue14}
}\begin{align}
\rho(z, p_z, t) = A \exp\left[ -a(t) z^2 - b(t) p_z^2 - c(t) {z} {p_z}\right]. \label{eq:bme_solution}
\end{align}
{This leads to a system of three coupled linear differential equations of first-order for the time-dependent coefficients $a(t)$, $b(t)$ and $c(t)$}\begin{align}
\dot{a} &= m \omega_z^2 c  \label{eq:dot_alpha} \\
\dot{b} &= - c / m \label{eq:dot_beta} \\
\dot{c} &= 2 m \omega_z^2 b - 2 a / m. \label{eq:dot_gamma}
\end{align}
{Equations~(7)-(10) can be used to compute  analytical expressions for the relative entropies (2) (Appendix B).} 

{The evolution of the phase-space density $\rho(z, p_z,t)$, together with the relative entropy $D_{zp_z}(t)= D(\rho(t) \vert\vert \rho_f)$ and  the positional relative entropy $D_z(t) = D(f_i(t) \vert \vert f_f)$ are shown in Fig.~\ref{fig:harmonic_case}a as a function of time. The phase-space distribution $\rho(z, p_z,t)$ is circular (equilibrium) for the initial  thermal state. It is elliptic (nonequilibrium) after the first Raman pulse applied at $t=0$ and rotates with period $t_\text{osc} = 2\pi /\omega_z$. It is again circular (equilibrium) for the thermal state attained after the second Raman pulse applied at $t = (3/4) t_\text{osc}$. The relative entropy $D_{zp_z}(t)$ is constant during free evolution. Its value is halved after each Raman pulse (dots) until it vanishes once the target state $\rho_f$ is reached. This is the effect captured by the first (static) thermalization criterion. On the other hand, the positional relative entropy $D_z(t)$ is only constant for equilibrium states and oscillates for nonequilibrium distributions, reflecting the rotation of the phase-space density $\rho(z, p_z,t)$. The amplitude of these oscillations vanishes once the target state $\rho_f$ is reached. This is the physical content of the second (dynamical) thermalization criterion. Note that we have applied the second Raman pulse at $t = (3/4) t_\text{osc}$ {in} this example only to display the oscillations of the intermediate nonequilibrium state. Optimal thermalization can already be achieved at  $t = (1/4) t_\text{osc}$.}
{We may further characterize the cooling process in one single diagram by combining the relevant quantities for the static and dynamical  criteria, $D_{zp_z}$ and $\Delta D$, (Fig.~\ref{fig:harmonic_case}b),  in a   schematic representation that qualitatively resembles Fig.~\ref{fig:overview}.}

{For the ideal harmonic trap, full phase-space thermalization at the Raman cooling temperature $\Tc$ is already obtained  after the second DRSC pulse.  The situation is more involved for nonharmonic  potentials. Owing to the nonlinearity of the trapping force, each atom has a different period which depends on the oscillation amplitude. Determining the optimal pulse spacing for arbitrary nonharmonic potentials is therefore a highly nontrivial task. We will next show that our optimization strategy successfully works for arbitrary potentials.}

\section{Experimental Setup}

\begin{figure}[t]
	\begin{center}
		\includegraphics[scale=1]{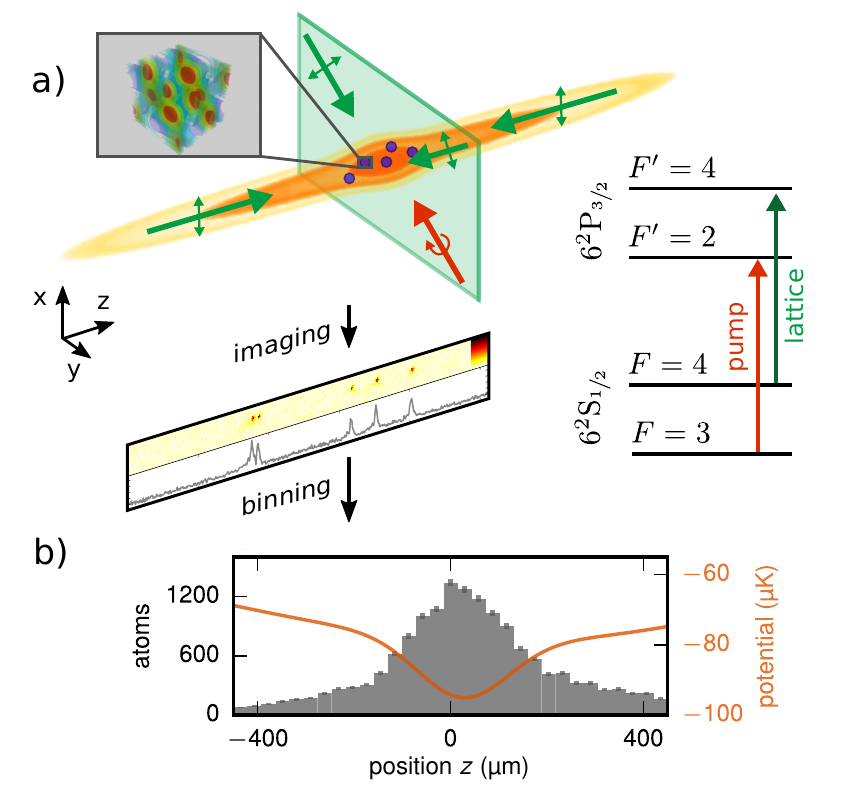}
	\end{center}
	\caption{Experimental setup and analysis.
		\textbf{a)} {Cs atoms (blue) are held in a crossed optical dipole trap (orange equipotential surfaces). In oder to realize the DRSC, four lattice beams (green arrows) and a pump beam (red arrow) are employed which are tuned close to the resonance of the Cs-D2 transitions indicated in the sketch.
		The polarizations of the beams are indicated by the corresponding small arrows.
		The interference of the four DRSC lattice laser beams creates the three-dimensional potential landscape illustrated in the inset.}
		The axial position distribution of the atoms is experimentally determined by employing fluorescence imaging in a one-dimensional optical lattice.
		\textbf{b)} {Typical measurement result, showing the} initial  position distribution (grey bars) with temperature $\Ti = \tempFOVEinit$ in the nonharmonic trapping potential (orange line).} 
	\label{fig:setup}
\end{figure}

We initialize our system by trapping  an average of 7 Cs atoms from background vapor in a  magneto-optical trap and transfer them into a crossed optical-dipole trap which creates a conservative potential (Fig.~\ref{fig:setup}a).
{The trap is formed by a horizontal laser beam propagating along the $z$-axis with a beam waist of $\SI{21}{\micro\meter}$ and power of $\SI{0.25}{\watt}$, and a second  crossed, vertical beam pointing in $-x$-direction with a waist of $\SI{165}{\micro\meter}$ and power of $\SI{3.5}{\watt}$.} The atomic collision rate of $\SI{36}{\hertz}$ at peak density is smaller than the inverse evolution time used in the experiment.
The cloud is thus effectively noninteracting. 
We extract the atomic positions along the axial $z$-direction by employing fluorescence imaging in a 1D optical lattice \cite{sch16} and obtain the experimental position distribution $f(z)$ after binning (Fig.~\ref{fig:setup}b). Every measurement is repeated several hundred times with identical parameters to get sufficient statistics.
The dipole trap potential  is approximately harmonic in radial direction $(x,y)$ with trap frequency $\omega_r = 2 \pi \times \SI{1.1}{\kilo\hertz}$.
The initial thermal state at temperature $T_0$ is prepared by applying a {sufficiently long} optical molasses pulse \cite{met99}.
The potential is markedly anharmonic in the axial $z$-direction and the position distribution features pronounced wings. At the center of the trap, the harmonic approximation yields an axial  frequency of $\omega_a = 2 \pi \times \SI{60}{\hertz}$.
We extract the initial temperature of the gas, $T_0 = \tempFOVEinit$, by comparing the measured position distributions $f(z)$ to numerical simulations of the three-dimensional trapping potential for atoms at various temperatures in a $\chi^2-$analysis (Appendix C). 

\begin{figure}[t]
	\begin{center}
		\includegraphics[scale=1]{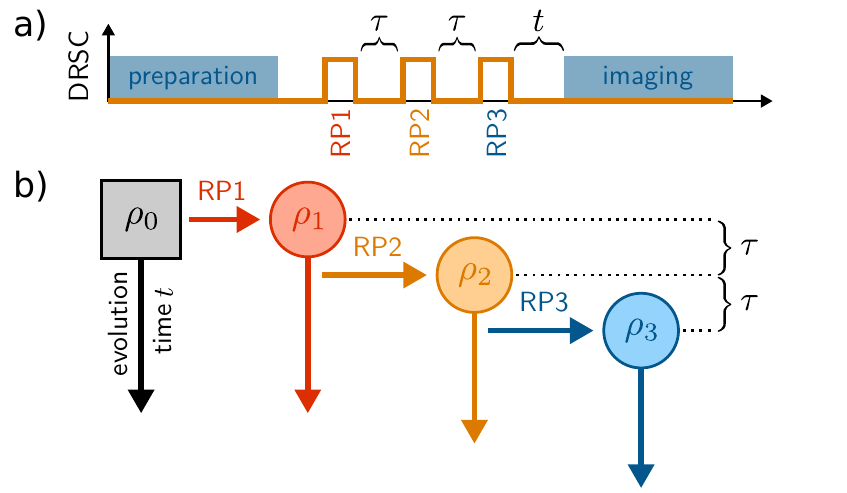}
	\end{center}
	\caption{Experimental sequence.
		\textbf{a)} After the initial preparation, three DRSC pulses RP1, RP2, and RP3 are applied with a pulse spacing of $\tau$.
		The final state is investigated by position resolved fluorescence imaging after an evolution time $t$.
		\textbf{b)} The Raman pulses convert the initial  state $\rho_0$ into the respective states $\rho_1$,  $\rho_2$,  and $\rho_3$.
		The time evolution of all the states is accessible experimentally by successively disabling the DRSC pulses shown in a) at the right stage.} 
	\label{fig:sequence}
\end{figure}

{We  cool the initial state of the atomic cloud by applying a train of DRSC pulses following the scheme of Ref.~\cite{ker00}. Details on the experiment may be found in   Refs.~\cite{hoh16, may18}.}
The setup comprises four DRSC lattice beams and a pump beam, as illustrated in Fig.~\ref{fig:setup}a.
At a detuning of $\SI{-6}{\mega\hertz}$ from the Cs $\mathrm{D}_2$-transition $\ket{F=4}\rightarrow\ket{F'=4}$, the DRSC lattice lasers create an interference pattern with lattice sites at a trap depth of $U_\mathrm{Raman} = k \times \SI{44}{\micro\kelvin}$ and trap frequencies of $\omega_\mathrm{trap} = 2 \pi \times (71, 29, 28) \,\si{\kilo \hertz}$ along the principal axis of the trap minimum.
During a DRSC pulse the Cs atoms are tightly confined in a lattice site.
The magnetic background field of $\SI{100}{\milli G}$ {applied along the $x$-$y$-diagonal} is chosen such that neighboring Zeeman and vibrational states $\ket{m_F+1, \nu}$ and $\ket{m_F, \nu-1}$ are energetically degenerate.
A Raman coupling induced by the DRSC lasers leads to the exchange of population between these degenerate states and thereby facilitates the transfer of vibrational energy to Zeeman energy.
An additional DRSC pumping beam which drives mainly $\sigma^+$-transitions at a detuning of $\SI{12}{\mega\hertz}$ to the $\ket{F=3}\rightarrow\ket{F'=2}$-transition dissipates the Zeeman-energy, while preserving the vibrational state during the absorption and subsequent emission of the pump-photons.
The Lamb-Dicke factors along the three principal axis of the Raman lattice sites are $\eta = (0.17, 0.27, 0.27)$.
This leads on average to a reduction of the vibrational quantum number $\nu$, and thereby to a decrease of the kinetic energy of the atoms.}

{We apply a train of three such \ramancooling{} pulses with duration of $\SI{10}{\milli \second}$ each and equal spacing of $\tau$ to the atomic sample as illustrated in Fig.~\ref{fig:sequence}a.
The  state $\rho_3$ resulting from this protocol is imaged after a variable evolution time $t$, which allows to record the time evolution of the state.
{Interrupting the DRSC protocol at  any intermediate  step $i$ by disabling subsequent pulses  provides experimental access to  the intermediate states $\rho_i$ (Fig.~\ref{fig:sequence}b).}

\section{Numerical Phase-Space Reconstruction}
\begin{figure*}
	\begin{center}
		\includegraphics[scale=1]{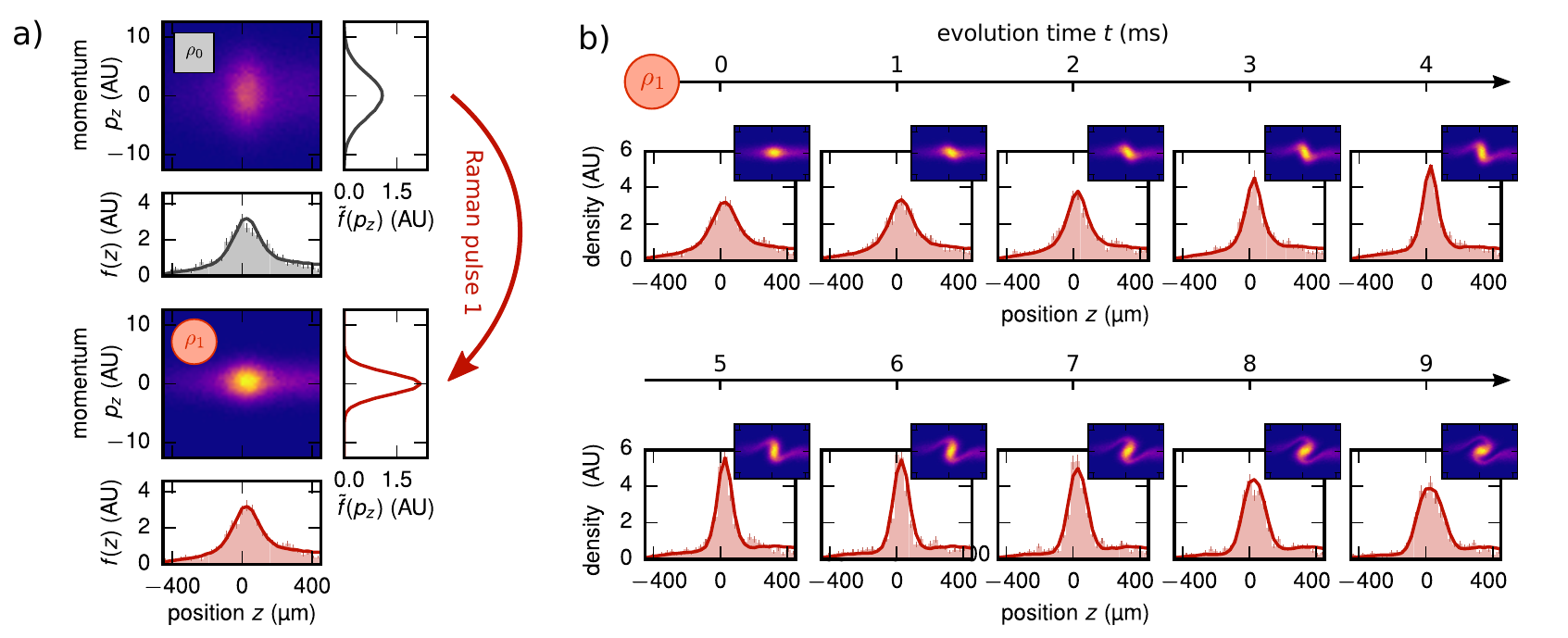}
	\end{center}
	\caption{DRSC characterization.
		\textbf{a)} Effect of the first Raman pulse on the axial phase-space distribution.
		The cooling effect acts on the momentum distribution only (red arrows in $\tilde{f}(p_z)$) and redistributes the momenta to a Maxwell-Boltzmann distribution corresponding to the Raman temperature $\Tc$.
		The initial thermal state $\rho_0$ is thereby driven into a nonequilibrium state $\rho_1$, with position and momentum distributions, $f_1(z)$ and $\tilde{f}_1(p_z)$, at a different temperature. This imbalance leads to the time evolution shown in b.
		\textbf{b)} Simulated free time evolution of the phase-space and corresponding projected axial density distribution (red solid line). The comparison with experimental data (red bars) allows us to determine the Raman temperature $\Tc = \tempFOVEcooled$.} 
	\label{fig:drsc_evolution}
\end{figure*}

{The specific properties of the DRSC interaction facilitate a simple effective description of the cooling effect.
First, the tight confinement of the Cs atoms in the 3D DRSC lattice potential pins the atomic position to a specific DRSC lattice site.
The lattice spacing of the DRSC lattice is of the order of $\SI{1}{\micro\meter}$, which is much less  than the typical dimension of the atomic sample in the optical dipole trap. The positions of the atoms in the dipole trap are therefore effectively frozen and  the position distribution $f(z)$ does not change  during the DRSC.
Second, the cooling effect of the DRSC imposes a new distribution of atomic momenta to the sample.
This distribution can be described in good approximation by a Maxwell-Boltzmann distribution. The temperature $\Tc$ that characterizes the momentum distribution $\tilde f(p_z)$ will be referred to as Raman temperature.}
Since the potential energy in the crossed dipole trap is unchanged, the DRSC pulse creates in general a nonthermal state.

{The validity of this effective description of the DRSC is confirmed by experimental data in Fig.~\ref{fig:drsc_evolution}a  for the first Raman cooling pulse. 
The measurements of the position distribution before and after the pulse verify that it remains unchanged during the DRSC pulse.
The effect of the DRSC in the momentum distribution can be studied by observing the free evolution of the system.
The state $\rho_1$ after the first pulse is not thermal since position and momentum distributions correspond to different temperatures, $\Ti$ and $\Tc$, respectively.
This imbalance gives rise to the phase-space dynamics shown  in Fig.~\ref{fig:drsc_evolution}b and can be employed to extract the Raman temperature $\Tc$.}
We compare the measured evolution of the position distribution $f_1(z)$ to numerical simulations of the three-dimensional trapping potential with the temperature $\Tc$ being the only free parameter. We obtain a Raman temperature of $\Tc = \tempFOVEcooled$ in  a $\chi^2$-analysis  (Appendix C).}
The simulation data can additionally be used {as an efficient way} to extract the full phase-space information as shown in the insets of Fig.~\ref{fig:drsc_evolution}b.
{This information is commonly only available at the price of additional technical effort or much larger atom number than used here~\cite{afe17, ber18}.}
While the axial phase-space distribution $\rho_1$ would simply  freely rotate in the two-dimensional space $(z,p_z)$ for a harmonic  potential, we here observe the creation of whorls induced by the nonlinearity of the trap \cite{mil86}. The projection onto the position axis shows excellent agreement between numerics and experimental data at all times. 
{We may thus conclude that the effective model for the DRSC captures all the relevant features of the phase-space evolution.
We can further simulate the full cooling protocol without any free parameters, once we have determined the initial and Raman temperatures.}

{The  properties of the DRSC also enable the evaluation of the relative entropy (2) right after a DRSC pulse (at evolution time $t=0$ in Fig.~\ref{fig:drsc_evolution}b).} Since the momentum distribution {is randomized to the same Maxwell distribution, $\tilde f_i(p_z)= \tilde f_f(p_z)$, characterized by only the Raman temperature $\Tc$ during each Raman pulse}, it is independent from the position distribution $f_i(z)$. As a result, the phase-space distribution factorizes $\rho_i(z,p_z)= f_i(z) \tilde f_f(p_z)$ directly after a Raman pulse. We can thus determine the full axial phase-space distribution  $\rho_i(z,p_z)$ immediately after each cooling pulse. Since the factorization property also holds true for a thermal state, we have for the final state $\rho_f(z,p_z) = f_f(z) \tilde f_f(p_z)$. The additivity of the relative entropy for independent distributions \cite{cov06} then implies that the entropic distance between $\rho_i$ and the target state $\rho_f$ simplifies to 
\begin{align}
D\left(\rho_{i}||\rho_f\right) &= D(f_i(z)||f_f(z))+D(\tilde f_f(p_z)||\tilde f_f(p_z)) \\
&= D(f_i(z)||f_f(z)).
\end{align}
{The full relative entropy  can hence be determined from the measured position distribution $f_i(z)$.} {We next discuss how this central quantity of nonequilibrium thermodynamics can be evaluated from experimental data in order to optimize the cooling of the few-particle  gas.}

\section{Application to Experimental Data}
\subsection{Optimal thermalization}

\begin{figure}[t]
	\begin{center}
		\includegraphics[scale=1]{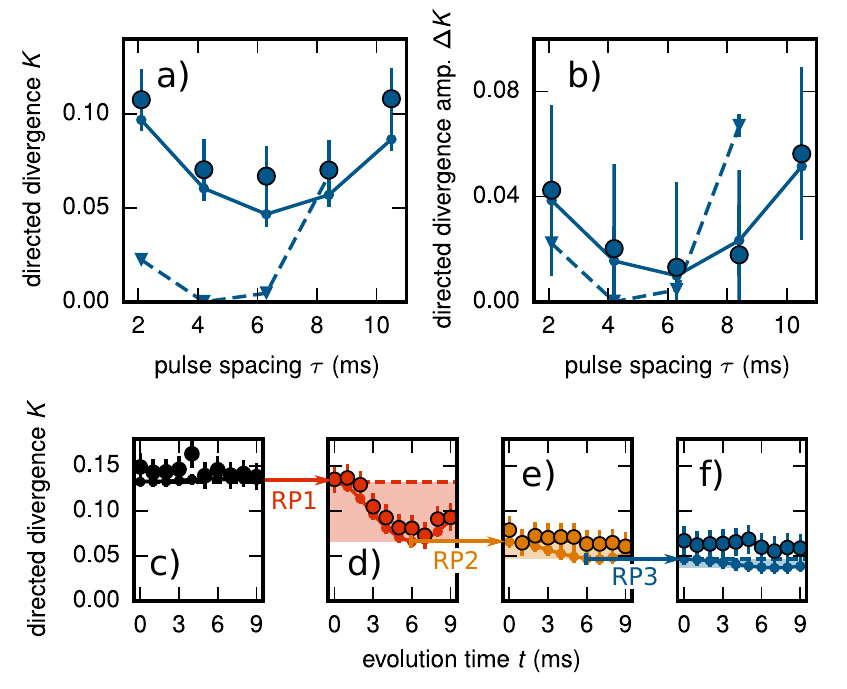}
	\end{center}
	\caption{Optimization criteria. \textbf{a)} Directed divergence $K\left(\rho_{3}||\rho_f\right)$ between the state $\rho_3$ after the last Raman pulse and the target thermal state $\rho_f$ for different pulse spacings: small triangles (dots) correspond to simulations of the harmonic (nonharmonic) trapping potentials and large dots show experimental data. \textbf{b)}  Oscillation amplitude $\Delta K = \max_t K(f_3(t)||f_f) - \min_t K(f_3(t)||f_f)$ for the position distribution $f_3$ after the last cooling pulse and the target position distribution $f_f$ for various pulse spacings. Both conditions yield an optimal spacing of $\SI{6.3}{\milli\second}$ for the nonharmonic experimental trap and $\SI{4.2}{\milli\second}$ for the harmonic trap. \textbf{c)-f)} Time evolution of {the position contribution to the directed divergence $K(f_3(t)||f_f)$} after each Raman pulse for the optimal time $\SI{6.3}{\milli\second}$: no oscillations are seen for the initial thermal state (black), while they increase after the first cooling pulse (red), before decreasing again after each Raman pulse that lead to thermalization (orange and blue).
	}
	\label{fig:optimization_pulsespacing}
\end{figure}

\begin{figure}[t]
	\begin{center}
		\includegraphics[scale=1]{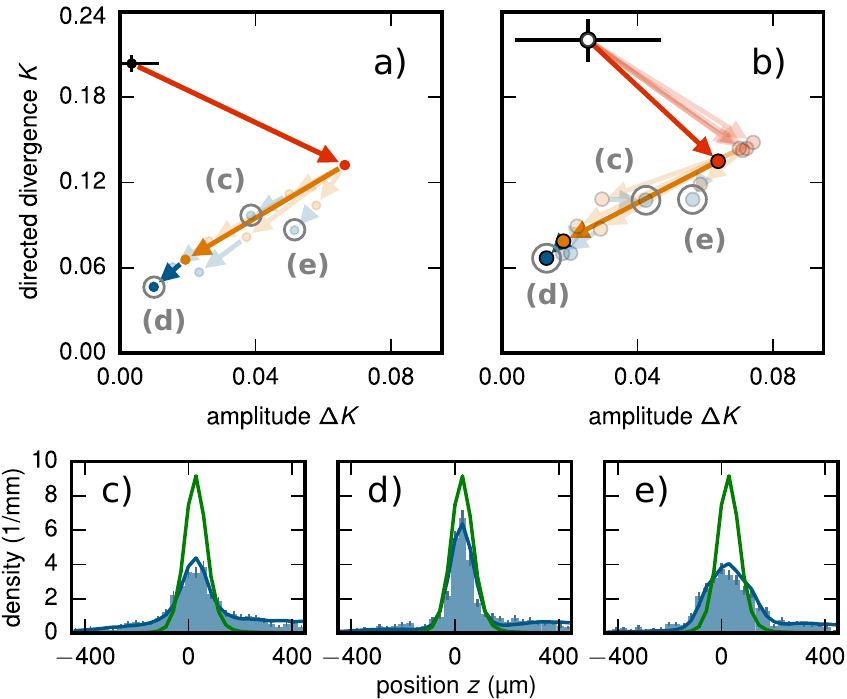}
	\end{center}
	\caption{
		Cooling map. \textbf{a)} Simulated cooling process in the plane $(K,\Delta K)$ of the two optimization quantities of Fig.~2. The initial thermal state  is shown in black.
		The red, yellow and blue arrows visualize the effect of the individual cooling pulses.
		The last states  for $\tau = $ \SIlist{2.1; 6.3; 10.5}{\milli \second} are labeled with (c), (d) and (e).
		\textbf{b)} Corresponding experimental cooling process. The hollow point includes the numerical contribution of the directed divergence of the momentum  distribution (Appendix E).
		\textbf{c)-e)} Atom distributions after the last Raman  pulse for $\tau = $ \SIlist{2.1;6.3;10.5}{\milli\second}.
		The experimental distribution (blue bars) and the corresponding simulation (blue solid line) are shown with the simulated target distribution (green solid line) as a reference. An overlap of $75\%$ is obtained for the optimal spacing d).
	} 
	\label{fig:optimization_map}
\end{figure}

The practical implementation of the two optimization criteria based on the total and positional relative entropies faces the problem that the relative entropy is only well-defined for probability distributions that are absolutely continuous with respect to one another, that is, there exists no point in phase space where one distribution vanishes, while the other one does not \cite{lin91}. Any occurrence of zero bins, due to finite statistics, in the experimentally measured or in the numerically simulated   distribution in the denominator will thus result, for a nonvanishing numerator, in a division by zero (Appendix D). {This issue does not seem to have been noticed in the nonequilibrium thermodynamics literature so far \cite{bus05,sek10,sei12,jar11,cil13}. We solve it by
replacing the relative entropy by the closely related $K$ directed divergence, well-known in engineering, and defined as \cite{lin91}}
\begin{align}
K(\rho_a||\rho_b) = D(\rho_a||(\rho_a+\rho_b)/2).
\end{align}
It satisfies $K(\rho_a||\rho_b)\geq 0$ and $K(\rho_a||\rho_b)=0$ if and only if $\rho_a = \rho_b$, like the relative entropy (2). It is always well defined irrespective of $\rho_a$ and $\rho_b$. It is further bounded by the relative entropy, $K(\rho_a||\rho_b) \leq D(\rho_a||\rho_b)/2$ \cite{lin91}. {It thus provides a lower bound to the energy irreversibly dissipated from the system during thermalization.} {We shall see {below} that the use of the $K$ directed divergence allows the optimal thermalization of the atomic gas.} 

\begin{figure}[t]
	\begin{center}
		\includegraphics[scale=1]{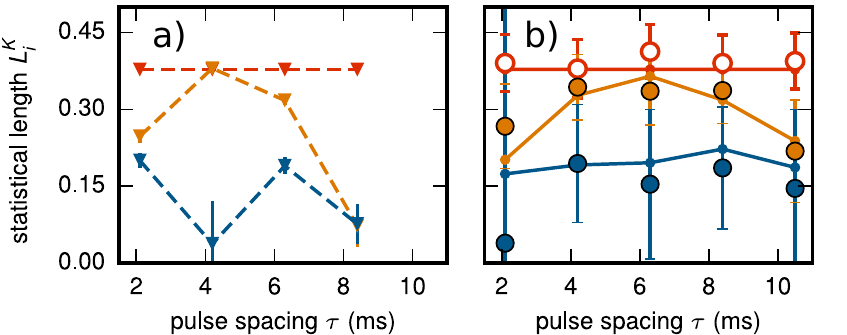}
	\end{center}
	\caption{
		Statistical length between cooling pulses.
		\textbf{a)} Simulated statistical length  $L_i^K$ from the $K$ directed divergence for $i=1, 2, 3$ (red, yellow, blue), in the harmonic trap for various spacings. \textbf{b)} Measured statistical length for the anharmonic trap (large points) and corresponding simulations. In both cases, thermalization is mainly reached during the first two steps for the optimal spacing, with nearly equal $L_i^K$. The hollow points include the numerical contribution of the directed divergence of the momentum  distribution (Appendix E).
	} 
	\label{fig:excess_entropy}
\end{figure}

Figure~\ref{fig:optimization_pulsespacing}a  presents the implementation of the first (static) optimization criterion for the state $\rho_3$. The $K$ directed divergence $K\left(\rho_{3}||\rho_f\right)$ is shown for various pulse spacings: the triangles correspond to numerical simulations for a harmonic trap, while the large dots are the experimental results for the nonharmonic trap. The small dots are the related simulations (Appendix D). We observe a vanishing minimum in the harmonic case  at  $\SI{4.2}{\milli\second}$ which corresponds to a quarter of a trap period. The state $\rho_3$ after the last Raman pulse is here  equal to the target thermal state $\rho_f$, revealing perfect thermalization.  We experimentally find a minimum  for the nonharmonic case at $\SI{6.3}{\milli\second}$, in good agreement with the numerical simulations. The entropic distance to the target state $\rho_f$ is reduced by almost a factor two at this point compared to the nonoptimal protocols.

Figure~\ref{fig:optimization_pulsespacing}b displays the results of the second (dynamical) optimization criterion for the state $\rho_3$. The oscillation amplitude $\Delta K = \max_t K(f_3(t)||f_f) - \min_t K(f_3(t)||f_f)$ after the last cooling pulse for a free evolution up to \SI{9}{\milli\second} is shown for different pulse spacings, both for the harmonic (triangles) and anharmonic (dots) potentials.  We again observe a minimum at $\SI{4.2}{\milli\second}$ for the simulated harmonic case and at $\SI{6.3}{\milli\second}$ for the experimental nonharmonic potential, thus confirming the findings obtained with the first, static condition. Figures~\ref{fig:optimization_pulsespacing}c-f show the time evolution of the $K$ directed divergence $K(f_3(t)||f_f)$  after each cooling pulse for the optimal spacing. No oscillations are seen for the initial thermal state $\rho_0$ (black). These oscillations   strongly increase after the first cooling pulse (red), revealing the nonthermal nature of state $\rho_1$, before decreasing again for the states $\rho_2$ and $\rho_3$ after the application of each additional Raman pulse (orange and blue). Finally, the oscillation amplitude reaches a minimum for $\rho_3$.

Both criteria may be combined to draw a map (Figs.~\ref{fig:optimization_map}a-b) of the cooling process in the plane {$(K\left(\rho_{i}||\rho_f\right),\Delta K(f_{i}))$}, similar to Fig.~\ref{fig:overview}a.  Figures~\ref{fig:optimization_map}c-d further show the overlap between the measured (blue bars) and simulated (blue lines) axial  distributions after the last pulse, as well as the simulated target distribution (green lines) for  $\tau = $ \SIlist{2.1;6.3;10.5}{\milli\second}. We observe an overlap of $75\%$ for the optimal spacing of $\SI{6.3}{\milli\second}$, twice the value for the other two times (Appendix F). {This offers an additional confirmation of the validity of the two thermodynamic optimization criteria. We note, however, that the overlap integral does not possess any simple thermodynamic interpretation in contrast to the relative entropy or the $K$ directed divergence.}

\begin{figure}[t]
	\begin{center}
		\includegraphics[scale=1]{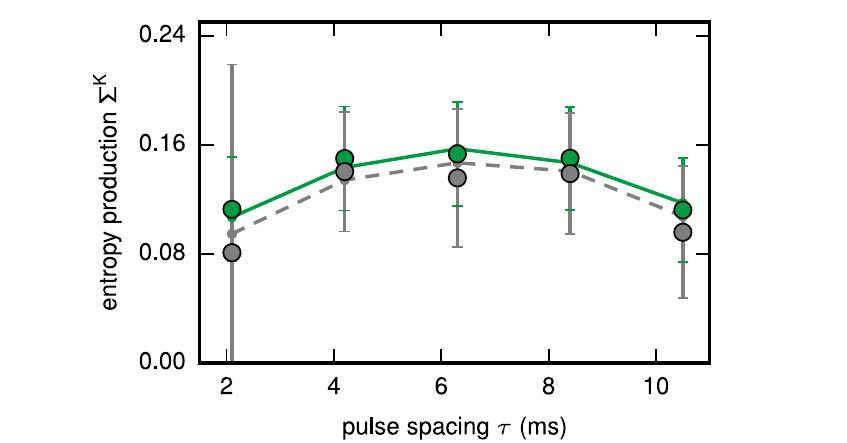} 
	\end{center}
	\caption{Verification of the generalized horse-carrot theorem, $\Sigma^K \geq (L^K)^2/(2n)$, for the $K$ directed divergence, showing that the total entropy production, $\Sigma^K = \sum_{i=1}^n \Sigma^K_i$, (green dots/data) (green solid lines/simulations) is bounded from below by the square of the total statistical length, $L^K= \sum_{i=1}^n L^K_i$, divided by twice the number of steps $n$ (grey dots/data) (grey dashed lines/simulations).	
	}
	\label{fig:horse_carrot}
\end{figure}

\subsection {Statistical length and  horse-carrot theorem}
The experimental reconstruction of the axial phase-space distribution after each Raman pulse allows us to analyze the whole cooling process by evaluating the statistical length $L_i$ of each cooling step. Figure~\ref{fig:excess_entropy}a presents the simulated lengths $L_i^{K}, (i=1, 2, 3)$ based on the $K$ directed divergence for the harmonic trap. We note that the first two steps (red and orange) have equal length for the optimal  spacing of  $\SI{4.2}{\milli\second}$, while the length of the last step vanishes. Optimal thermalization thus occurs during the first two Raman pulses with identical entropy production. This picture is still approximately true for the nonharmonic potential (Fig.~\ref{fig:excess_entropy}b): the first two statistical lengths are nearly equal for the optimal pulse spacing of $\SI{6.3}{\milli\second}$, while the third one is much smaller. {This is a nontrivial  result: it was originally theoretically derived for close-to-equilibrium quasistatic processes \cite{sal83,and84,nul85,and11,sal98,sal01,nul02} (see also Refs.~\cite{ton87,spi95,dio96}) and has never been confirmed experimentally to our knowledge. The fact that it also holds true (exactly for the harmonic case and approximately for the anharmonic trap) for the generalized statistical length (4) (even when the relative entropy is replaced by the $K$ directed divergence) is remarkable. It suggests a quite general range of validity of the principle of equal statistical distances (or of equipartition of entropy production, as it is sometimes called \cite{ton87,spi95,dio96}) for optimal nonequilibrium  processes. Figure~\ref{fig:horse_carrot} additionally shows an experimental verification of the generalized horse-carrot theorem (5),  $\Sigma^K \geq (L^K)^2/(2n)$ for the $K$ directed divergence, as a function of the pulse spacing. It shows that the entropy production is maximal for the optimal pulse spacing, corresponding to maximal heat extraction from the system.  This situation is somewhat different from the usual one, where the system of interest is continuously coupled to an ideal heat bath. In the present experiment, the phase-space evolution is mostly nondissipative as the system is only punctually coupled to a nonideal reservoir that thermalize the momentum degree of freedom. Figure~\ref{fig:horse_carrot} confirms the validity of a sharpened second law in this nonequilibrium situation.}

\section{Conclusion}
{We have experimentally studied the nonequilibrium  thermodynamics  of  a few-particle system consisting of a gas of  noninteracting Cesium atoms driven by Raman laser cooling pulses. Tracing the evolution of the gas with position-resolved fluorescence imaging enabled us to access the full phase-space density of the effectively one-dimensional system. We have used this distribution to evaluate the nonequilibrium entropy production and the statistical length based on the $K$ directed divergence. The latter quantity is always defined, in contrast to the usual relative entropy, and provides a lower bound to it. It  further belongs to the family of $f$-divergences and shares their properties \cite{csi67}. As a first application, we have optimized the thermalization of the atomic gas and determined the optimal Raman pulse spacing for a nonharmonic trap potential by minimizing the entropy production to a final target state.  We have additionally verified a horse-carrot theorem and analyzed the entire cooling process with the help of the statistical length. We have found that optimal thermalization is mainly achieved during the first two cooling stages, corresponding to nearly equal statistical lengths.
{Our findings demonstrate an effective, theoretical and experimental, method to characterize and optimize general nonequilibrium processes of few-particle systems.} They further highlight the practical usefulness of nonequilibrium concepts such as entropy production and statistical lengths down to the atomic level. {While we have validated our generic approach with the example of  laser cooling of  noninteracting atoms, the same theoretical and experimental techniques can be straightforwardly employed to include external time-dependent drivings, tunable interactions or dissipation effects. Our results thus provide a versatile platform to engineer nonequilibrium states and investigate complex far-from-equilibrium-optimization protocols for driven-dissipative interacting particles \cite{lab16}, as well as for power output mechanisms and thermal machines \cite{ros16}, both in the classical and quantum regimes.}

\section*{Acknowledgements}
We acknowledge  financial support from the German Science Foundation (DFG) under Project No.~277625399 - TRR 185 and Grant No.~FOR 2724.

\section*{Appendix A: Entropy production and statistical length}
We begin by reminding the derivation of the entropy production for a single equilibration step \cite{def11}. We consider a system with Hamiltonian $H$ in an initial state $\rho_0$ that thermalizes to the equilibrium state $\rho_\text{eq}$ with inverse temperature $\beta$. The entropy production is defined as $\Sigma = \Delta S -\beta Q$, where $\Delta S = - \int dzdp_z\,(\rho_\text{eq}\ln \rho_\text{eq} - \rho_0\ln \rho_0)$ is the entropy difference between final and initial states and $Q=\int dzdp_z\, H(\rho_\text{eq}- \rho_0)$  the corresponding heat. Using $\rho_\text{eq} = \exp(-\beta H)/Z$, one readily finds \cite{sch80,pro76,def11},
\begin{equation}
\label{2}
\Sigma = D(\rho_0||\rho_\text{eq}) = \int dzdp_z\,(\rho_0\ln \rho_0 - \rho_0\ln \rho_\text{eq}).
\end{equation}
{Expression \eqref{2} is the maximal amount of work that can be extracted  during thermalization  \cite{pro76,def11}.}
Let us now consider a multistep equilibration process with one intermediate (nonthermal) state $\rho_1$. The entropy production between this state and the equilibrium state $\rho_\text{eq}$ is  $\Sigma_1 = D(\rho_1||\rho_\text{eq})$. Using the additivity of the entropy production, $\Sigma= \Sigma _0+ \Sigma_1$, we obtain the entropy production between state $\rho_0$ and $\rho_1$ as \cite{esp10,def11} (see also Refs.~\cite{cus18,man18}),
\begin{equation}
\Sigma_0 = D(\rho_0||\rho_\text{eq}) -D(\rho_1||\rho_\text{eq}).
\label{4}
\end{equation}
 Equation \eqref{4} can be generalized to an arbitrary number of nonthermal intermediate steps by recursion, yielding,
\begin{equation}
\Sigma_i = D(\rho_i||\rho_\text{eq}) -D(\rho_{i+1}||\rho_\text{eq}).
\label{5}
\end{equation}
{Common optimization schemes consider equilibrium intermediate states generated by coupling the system to different baths at (slightly) different temperatures $T_i$ \cite{nul85,and11,sal98,sal01,nul02}. In this quasistatic case, $\Sigma_i^\text{qs} = D(\rho_i||\rho_{\text{eq},i})$, where $\rho_{\text{eq},i}$ is a thermal state at inverse temperature $\beta_i$.  The square root, $L_i^\text{qs} =\sqrt{2 \Sigma_i^\text{qs}}$, defines a statistical length in thermodynamic space \cite{nul85,and11,sal98,sal01,nul02}. It is a proper (Riemannian) distance in contrast to the relative entropy that does not satisfy the triangle inequality. Interestingly, the total entropy production, $\Sigma^\text{qs} = \sum_{i=0}^n \Sigma_i^\text{qs}$, is bounded from below by the square of the total length $L^\text{qs}= \sum_{i=0}^n L_i^\text{qs}$, that is, $\Sigma^\text{qs} \geq (L^\text{qs})^2/(2n)$. This result, which follows from the Cauchy-Schwarz inequality, is often referred to as the horse-carrot theorem \cite{and11,sal98}. It is significant because it provides a sharper lower bound to the entropy production than the second law of thermodynamics, which only states that the entropy production is non-negative. The lower bound can actually be reached, showing that dissipation can be  reduced by coaxing the system along the desired path, much like guiding a horse along by waving a carrot in front of it \cite{and11,sal98}.}

{Similarly, the square root of Eq.~\eqref{5}, $L_i =\sqrt{2 \Sigma_i}$, defines a statistical length, which   reduces to the usual thermodynamic length for quasistatic processes \cite{nul85,and11,sal98,sal01,nul02}. The total entropy production is still bounded from below by the square of the total statistical length divided by twice the number of steps, $\Sigma \geq L^2/(2n)$, generalizing the horse-carrot theorem to nonthermal intermediate states. }

	\section*{Appendix B: Analytical Solution in the Harmonic Case}
	{Using  the analytical expression of the Gaussian phase-space density given in the main text, the position and momentum projections are easily integrated to 
	\begin{align*}
	f(z) = \int \rho(z, p_z) \d p_z &= \sqrt{\frac{a^*}{\pi}} \exp\left( -a^* z^2 \right) \\
	\tilde{f}(p_z) = \int \rho(z, p_z) \d z &= \sqrt{\frac{b^*}{\pi}} \exp\left( -b^* p_z^2 \right),
	\end{align*}
	where the time-dependent parameters $a^* =  a - {c^2}/{4b}$ and $b^* =  b - {c^2}/{4a}$ are the corresponding projected variables.
	The relative entropies follow as,
	\begin{align}
	D_{zp_z}(\rho_1||\rho_2) &= \frac{1}{2} \ln\left(\frac{4a_1 b_1 - c_1^2}{4a_2 b_2 - c_2^2}\right)  -1 \nonumber\\
	& + \frac{2 a_1 b_2 + 2 a_2 b_1 - c_1 c_2}{(4a_1 b_1 - c_1^2)^2} \\[3mm]
	D_z(f_1||f_2) & = \frac{1}{2} \left[\ln\left(\frac{a_1^*}{a_2^*}\right) + \frac{a_2^*}{a_1^*} - 1 \right]  \label{eq:position_relent_definition}	\end{align}
	The above expressions are employed for the calculations presented in Fig.~\ref{fig:harmonic_case} of the main text.
	The effect of the DRSC pulse is taken into account  by setting the parameters $a_1$, $b_1$, and $c_1$ before the DRSC pulse to new values $a_2$, $b_2$, and $c_2$, determined
	by incorporating the constrains arising from the characteristics of the cooling:
	First, the DRSC erases all correlations of the state, implying  $c_2=0$.
	Second, the velocity distribution is given by a Maxwellian  at the Raman cooling temperature $\Tc$.
	And third, the position distribution is not influenced by the DRSC.
	Summing up these conditions yields the parameters after the Raman cooling pulse to be
	\begin{align}
	a_2 &= a_1 - \frac{c_1^2}{4b_1} \\
	b_2 &= \frac{m}{2 k T_\mathrm{R}} \\
	c_2 &= 0.
	\end{align}
	}

\section*{Appendix C: Numerical Simulations}

For the numerical simulation of the phase-space dynamics in the \ramancooling{} protocols, the atomic motion in the trap is modeled with a Monte-Carlo approach where full three-dimensional trajectories of $N=10^5$ atoms are calculated.
This simulation only features two free parameters:
First, the initial temperature of the atomic cloud $\Ti$ determines the initial, thermal phase-space distribution, which sets the starting point for the simulation.
Second, the Raman cooling temperature $\Tc$ is employed to model the effect of the \ramancooling{} by resetting the atomic velocities to a Maxwell-Boltzmann distribution corresponding to $\Tc$, whenever a \ramancooling{} pulse is applied.
Using these two temperatures together with precise information on the trap, which was specified by independent trap frequency {and beam shape} measurements, the effect of arbitrary pulse sequences on the phase-space distribution and the ensuing dynamics can be computed.
In this section, we show how the experimental value for $\Ti$ is extracted from the measured initial distribution $f_0(z)$ and the Raman cooling temperature $\Tc$ is determined from the measured evolution after the first Raman cooling pulse $f_1(t, z)$.

\begin{figure}[t]
	\begin{center}
		\includegraphics[scale=1]{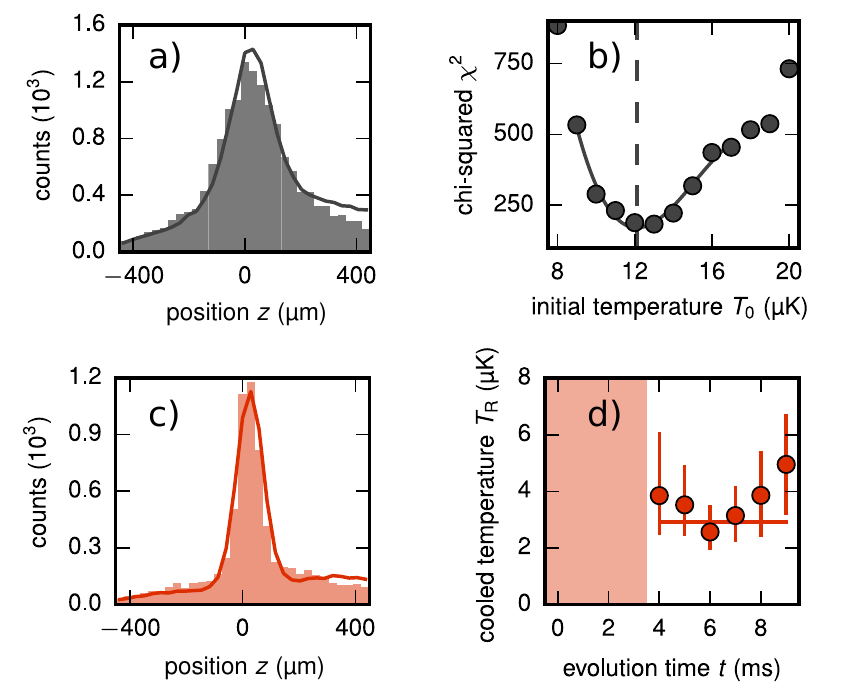}
	\end{center}
	\caption{
		Fitting of initial and final temperatures.
		a) Experimental position distribution (bars) and best fitting simulated distribution with $\Ti=\SI{12}{\micro\kelvin}$ (solid line).
		b) The initial temperature is extracted by calculating the $\chi^2-$value for various initial temperatures (markers) and then using a polynomial fit around the minimum of the curve (solid line) to extract the best fitting temperature $\Ti=\tempFOVEinit$ (dashed line).
		c) Experimental position distribution after a single \ramancooling{} pulse and $t=\SI{6}{\milli\second}$ evolution time (bars) and best fitting simulated distribution with $\Tc=\SI{3}{\micro\kelvin}$ (solid line).
		d) Applying a $\chi^2-$analysis for every evolution time $t$ yields different cooled temperature estimates (markers). We extract the overall cooled temperature $\Tc=\tempFOVEcooled$ by employing a weighted fit (solid line).
	}
	\label{fig:temperature_extraction}
\end{figure}

In order to model the position distribution $f_0(z)$, we employ a simulation scenario, where atoms are initially located at the trap center. 
A heat bath at temperature $\Ti$ is emulated by resetting the atomic velocities repeatedly to random velocities corresponding to the desired initial temperature $\Ti$.
Due to the resulting damped motion of the atoms in the trap, the atomic position distribution approaches a thermal distribution at $\Ti$ \cite{hoh16}.
We compare the simulated position distributions $f_\mathrm{sim}(z)$ for various temperatures $\Ti$ to the experimentally measured initial position distribution $f_\mathrm{exp}(z)$ shown in Fig.~\ref{fig:temperature_extraction}a by calculating the $\chi^2-$value,
\begin{eqnarray}
\chi^2 = \sum_{z_i} \left( \frac{f_\mathrm{sim}(z_i) - f_\mathrm{exp}(z_i)}{\Delta f_\mathrm{sim}(z_i) + \Delta f_\mathrm{exp}(z_i)} \right)^2
\end{eqnarray}
for the binned data as a measure for the goodness of the fit \cite{bev03} ($\Delta f_\mathrm{sim}$ and $\Delta f_\mathrm{exp}$ are the statistical uncertainties of $f_\mathrm{sim}$ and $f_\mathrm{exp}$).
The $\chi^2-$value for simulations at various temperatures is shown in Fig.~\ref{fig:temperature_extraction}b, where we use a polynomial fit to the data in order to extract the initial temperature $\Ti = \tempFOVEinit$.

The final temperature $\Tc$ which corresponds to the \ramancooling{} temperature is not visible in the position distribution directly after a \ramancooling{}-pulse.
However, as illustrated in Fig.~\ref{fig:drsc_evolution}b of the main text, the evolution in the trapping potential after the first \ramancooling{} pulse shows clear evidence of the cooling effect by featuring a breathing behavior.
In order to extract the value of $\Tc$, we simulate the time evolution of atomic samples which are prepared at the initial temperature $\Ti$ and then reset the atomic velocities to values corresponding to different Raman cooling temperatures $\Tc$.
For every evolution time $t$, we extract a Raman cooling temperature  $\Tc$ with a $\chi^2-$analysis, analogous to the strategy employed for the initial distribution,  by comparing the simulations for different Raman cooling temperatures to the experimental distribution (Fig.~\ref{fig:temperature_extraction}c).
We combine the results of all measured evolution times shown in Fig.~\ref{fig:temperature_extraction}d by a weighted constant fit to the data, thereby extracting the \ramancooling{} temperature $\Tc = \tempFOVEcooled$.
The red shaded area in the plot indicates small evolution times $t$ where the  $\chi^2-$analysis fails, because the information about the velocity distribution is not yet transformed into the position distribution.
This behavior is also visible in the size of the error bars, which first decreases until $t = \SI{6}{\milli\second}$ and then increases again.
{Combining the extracted values for $\Ti$ and $\Tc$, the simulation dataset corresponding to $\Ti = \SI{12}{\micro\kelvin}$ and $\Tc = \SI{3}{\micro\kelvin}$ is the best fitting simulation.
	Therefore, this dataset is employed for the calculation of the simulation data points presented in the main text.
	Accordingly, the final thermal state $\rho$ is also represented by the simulation data for a thermal state at temperature $\Tc = \SI{3}{\micro\kelvin}$.}
	
	{DRSC  is in general a subrecoil cooling scheme, because it fundamentally allows to reach subrecoil temperatures. The temperature of  $\SI{2.9}{\micro\kelvin}$ observed in the experiment is clearly above the recoil temperature of Cs which is  $\SI{0.1}{\micro\kelvin}$ for the DRSC laser light. This optimum is not reached due to technical limitations like laser power noise, laser linewidths, off-resonant photon scattering.}

\begin{figure}[t]
	\begin{center}
		\includegraphics[scale=1]{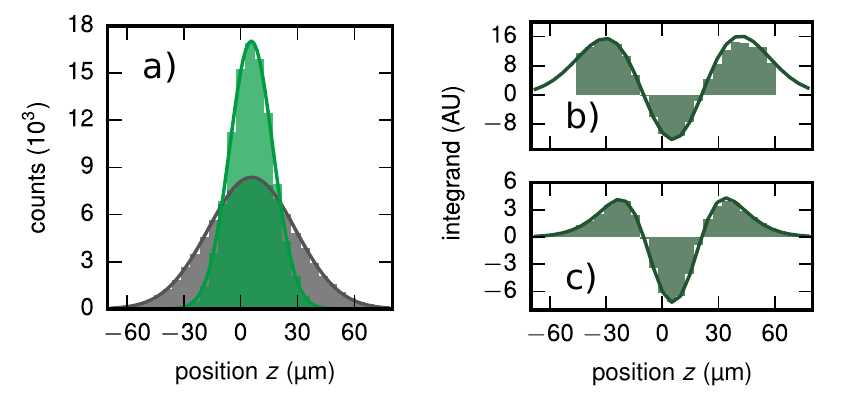} 
	\end{center}
	\caption{
		Numerical calculation of the relative entropy.
		a) Monte-Carlo simulations for an initial distribution at $\Ti=\SI{1}{\micro\kelvin}$ (gray bins) and a final distribution at $\Tc=\SI{0.25}{\micro\kelvin}$ (green bins).
		At these temperatures, the trapping potential is approximately harmonic,  Gaussian distributions (solid lines) can thus be fitted to the Monte-Carlo results.
		b) Integrand $\mathcal{D}_i$ for the calculation of the relative entropy $D(f_{i}||f_{f})$.
		c) Integrand  $\mathcal{K}_i$ for the calculation of the directed divergence $K(f_{i}||f_{f})$.
		Bins correspond to the calculation based on the  Monte-Carlo data, while solid lines represent the integrands based on the fitted Gaussian distributions shown in a).
	}
	\label{fig:num_rel_ent_harm}
\end{figure}

\section*{Appendix D: Numerical calculation of the relative entropy}

For the analysis of our data, we typically bin the atomic positions from an experiment or a Monte-Carlo simulation in order to create a numerical representation of the density distribution.
The integral for the relative entropy then corresponds to a sum over all bins $z_i$, where
\begin{eqnarray}
D(f_1 || f_2) &=& \sum_{z_i} f_1(z_i) \log\left(\frac{f_1(z_i)}{f_2(z_i)}\right) \delta z_i \label{eq:num_rel_ent} \\
&=& \sum_{z_i} \mathcal{D}_i \delta z_i.
\end{eqnarray}
As discussed in Ref.~\cite{lin91},
typical data with finite statistics may exhibit bins where $f_2(z_i)=0$, meaning that no atom has been observed in this bin.
However, this corresponds to a division by zero in Eq.~\eqref{eq:num_rel_ent}, rendering the calculation of the integrand value $\mathcal{D}_i$ impossible for this specific bin.
In contrast, the directed divergence \cite{lin91},
\begin{eqnarray}
K(f_1 || f_2) &=& \sum_{z_i} f_1(z_i) \log\left(\frac{f_1(z_i)}{f_1(z_i) + f_2(z_i)}\right) \delta z_i \label{eq:num_rel_en} \\
&=& \sum_{z_i} \mathcal{K}_i \delta z_i,
\end{eqnarray}
can be evaluated even at bins where $f_2(z_i)=0$. It is therefore  much more robust especially when analyzing experimental data, where statistical errors are usually even more pronounced.

In order to illustrate the problem, we employ the data set used for the harmonic approximations shown in Figs.~\ref{fig:optimization_pulsespacing} and \ref{fig:excess_entropy} of the main text. The corresponding initial temperature for the simulation is $\Ti=\SI{1}{\micro\kelvin}$ and the final temperature is $\Tc=\SI{0.25}{\micro\kelvin}$.
While these values are more than one order of magnitude colder than the experimental parameters, the ratio of the two temperatures is the same as in the experiment, thereby providing a comparable cooling process.
Nevertheless, at these low temperatures, the harmonic approximation of the trapping potential holds also in axial direction.
In fact, the density distributions of the Monte-Carlo simulation (bars) shown in Fig.~\ref{fig:num_rel_ent_harm}a fit very well to Gaussian distributions (solid lines) which are expected for the harmonic case.
Figures~\ref{fig:num_rel_ent_harm}b and c show the integrands $ \mathcal{K}_i$ and $\mathcal{D}_i$, where again the bars correspond to the numerical data and the solid lines show the Gaussian fit.
The missing bars seen in Fig.~\ref{fig:num_rel_ent_harm}b clearly indicate the numerical problem connected to the relative entropy. By contrast, the integral for the directed divergence in Fig.~\ref{fig:num_rel_ent_harm}c can be evaluated in the whole range.

\begin{figure}[t]
	\begin{center}
		\includegraphics[scale=1]{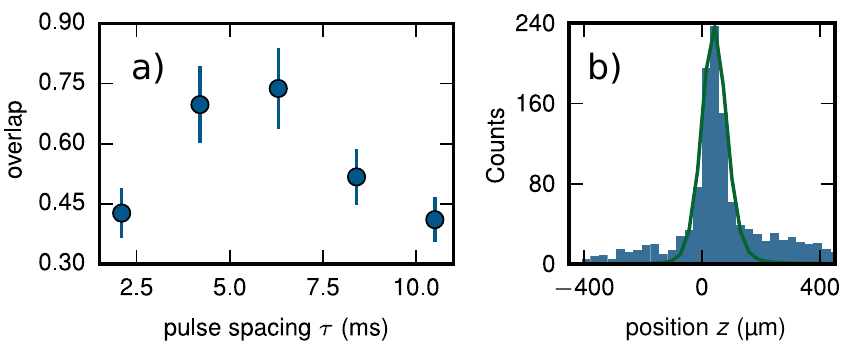} 
	\end{center}
	\caption{
		Overlap calculation.
		a) The overlap of the distribution after the last cooling pulse $f_3(0)$ with the final distribution $f_{f}$ shows a maximum at \SI{6.3}{\milli \second} in agreement with our optimization result.
		b) Illustration of the experimental distribution $f_3(0)$ for the optimum pulse spacing (blue bars) and the renormalized final distribution (green line).
	} 
	\label{fig:overlap}
\end{figure}

\section*{Appendix E: Contributions of the momentum distribution}
For factorized distributions $\rho_i = f_i(z) \tilde f_i(p_z)$ the relative entropy $D$ can be split into two contributions
$D\left(\rho_{i}||\rho_f\right) = D(f_i(z)||f_f(z))+D(\tilde f_i(p_z)||\tilde f_f(p_z))$, 
where the first term accounts for the position distribution and the second takes into account the momentum distribution.
After a Raman cooling pulse (for $i = 1, 2, 3$), the contribution of the momentum distributions is zero, because $f_i(p_z)$ and $f_f(p_z)$ are identical.
For the initial distribution ($i = 0$), however, this contribution is not zero, as here the momentum distributions are not equal.
In the measured position distributions $f_0(p_z)$ at $t=0$, this contribution is not visible.
However, as the initial ($\Ti$) and final ($\Tc$) temperatures are known, the contribution $D(\tilde f_0(p_z)||\tilde f_f(p_z))$ can be calculated from the thermal momentum distributions $\tilde f(p_z, T) = 1 / \sqrt{2 \pi m k T} \cdot \exp(- p_z^2 / (2mkT))$.
We find for the directed divergence employed in Figs.~\ref{fig:optimization_pulsespacing} and \ref{fig:optimization_map} this contribution of the momentum distribution to be $K(\tilde f(p_z, \Ti)||\tilde f(p_z, \Tc)) = 0.071$ by solving the integral numerically.
The hollow experimental points in Figs.~\ref{fig:optimization_pulsespacing} and \ref{fig:optimization_map} are thus a combination of the measured contribution to the directed divergence from the position distribution and the numerically deduced contribution from the momentum distribution.\\

\section*{Appendix F: Overlap calculation}

The  overlap of the distribution after the last cooling pulse $f_3(0)$ with the final distribution $f_{f}$ is evaluated in the following way. We first renormalize the final distribution (green line) to the maximum of the experimental data (blue bars).
Integration of the renormalized final distribution then yields the amount of atoms in the experimental distribution that match the final distribution.
We identify this value with the overlap. We find the largest overlap at a pulse spacing of \SI{6.3}{\milli \second} which corresponds to our optimization result  (Fig.~\ref{fig:overlap}).

\bibliographystyle{apsrev4-1}

\end{document}